\documentclass[review]{elsarticle}

\usepackage{amsmath,amsfonts,amssymb,amsthm,amstext}
\usepackage{bm}
\usepackage{dsfont}
\usepackage{braket}
\usepackage{graphicx}
\usepackage{lineno}

\journal{Current Opinion in Colloid \& Interface Science}

\begin{document}

\begin{frontmatter}

\title{Magnetic active matter across scales}

\author[chile]{Francisca Guzm\'an-Lastra\corref{cor1}}
\ead{fguzman@uchile.cl}
\address[chile]{Departamento de F\'isica, Facultad de Ciencias, Universidad de Chile, Santiago, Chile}

\author[hhu]{Margaret Rosenberg}
\ead{rosenberg@thphy.uni-duesseldorf.de}

\author[hhu]{Marco Musacchio}
\ead{marco.musacchio@hhu.de}

\author[rome]{Lorenzo Caprini}
\ead{lorenzo.caprini@uniroma1.it}

\author[hhu]{Hartmut L\"owen}
\ead{hlowen@hhu.de}

\address[hhu]{Institut f\"ur Theoretische Physik II: Weiche Materie, Heinrich-Heine-Universit\"at D\"usseldorf, Universit\"atsstra\ss e 1, D-40225 D\"usseldorf, Germany}
\address[rome]{Physics Department, University of Rome La Sapienza, P.le Aldo Moro 5, IT-00185 Rome, Italy}

\cortext[cor1]{Corresponding author}

\begin{abstract}
Magnetic interactions provide a versatile and powerful tool for controlling and organizing active matter, where individual units continuously consume energy to drive autonomous motion. These interactions arise naturally in biological systems, such as magnetotactic bacteria, and can be engineered into synthetic platforms, including colloidal microswimmers, magnetic nanoparticles, and macroscopic granular robots. This review focuses on active, self-propelled particles that carry an intrinsic magnetic dipole moment, powered by their own energy consumption rather than driven by external fields; here, the dipole moment mediates interactions and self-organization, not propulsion. We survey experimental and theoretical studies across all length scales, showing how dipolar interactions shape single-particle dynamics, collective behavior, and self-organization. We discuss models incorporating pairwise dipolar forces and confinement, and examine emergent phenomena such as chaining, swarming, and tunable pattern formation. We close by outlining challenges and opportunities in the design, control, and application of magnetic active systems, from programmable materials and biomedical actuation to nonequilibrium physics.
\end{abstract}


\end{frontmatter}

\section{Introduction}

Magnetic interactions have fascinated humans since antiquity, when lodestones were observed to attract iron. Early systematic studies by William Gilbert in De Magnete established that these forces arise from intrinsic material properties.
Today, magnetic interactions---most notably dipole--dipole forces---are recognized as fundamental mechanisms arising in a wide range of scientific and technological contexts, spanning condensed matter, colloidal, and biological systems~\cite{xie2019reconfigurable}. In magnetic materials, they contribute to frustration and to the emergence of fractionalized excitations in spin-ice systems~\cite{ortiz2019colloquium}, while also shaping domain structures in thin films and bulk ferromagnets. In soft-matter systems such as ferrofluids and magnetic colloids, they promote the formation of chains and other anisotropic assemblies under applied fields~\cite{kim2022magnetic}. Similar interactions also play a role in biological contexts, influencing the collective dynamics of magnetotactic microorganisms~\cite{jin2021collective}. 

Building on this broad physical relevance, magnetic interactions play an even richer role in active matter, where individual units continuously consume energy from the environment to sustain motion. Active matter is ubiquitous in nature since spans a wide range of length scales, from nanometer-scale molecular motor assemblies inside cells to meter-scale animals, including insects, fish and birds. At the micron scale, biological microswimmers, such as bacteria, spermatozoa and cells, provide prototypical examples of active systems. Beyond living organisms, synthetic active matter includes colloidal swimmers, magnetic nanoparticles~\cite{rigoni2025activerosensweigpatterns}, and macroscopic active granular systems~\cite{kumar2014flocking,casiulis2025geometric,antonov2024inertial} such as vibrobots~\cite{scholz2018rotating}, bristlebots~\cite{Reiche2025} or hexbug particles~\cite{obreque2026dynamics}, which exhibit self-propelled motion and collective behaviors analogous to those seen in biological systems.

Magnetic interactions are particularly relevant in active matter for two complementary reasons. First, they naturally arise in many biological systems, where internal magnetic moments---such as the chains of magnetic nanoparticles in magnetotactic bacteria---mediate alignment, steering, and collective motion. Second, these interactions can be readily incorporated into synthetic active systems, including colloidal swimmers, magnetic nanoparticles, and macroscopic granular robots, enabling precise control over propulsion, assembly, and emergent patterns~\cite{Waitukaitis25,Stikuts2026}. Together, these features make magnetic interactions a powerful and versatile tool for both probing fundamental principles of non-equilibrium organization and designing controllable active materials.

Tuning magnetic interactions in active matter may represent a promising route to address some challenges currently faced by our society. 
Magnetic active matter may serve to design biologically inspired materials capable of spontaneous self-assembly. More specifically, magnetic interactions may serve to shape and control collective phenomena, promoting programmable self-assembly and long-range orientational alignment~\cite{koessel2020emergent, martinez2015colloidal,kiani2015elastic,tierno2012steering}. 
One of the great challenges of active matter is related to medical applications, such as drug delivery in our body. Active micro- or nano-machines and nanomachines may perform tasks in an autonomous way. Magnetic interactions offer a promising platform for designing, using, and controlling microswimmers and nanoswimmers potentially delivering cargoes, through directional guidance or tunable collective behavior~\cite{thery2020selforganisation,vincenti2019magnetotactic},

In this review, we systematically survey experimental examples of active systems governed by magnetic dipole--dipole interactions across all length scales, encompassing both biological and synthetic systems (Sec.~\ref{Sec:2}). We further discuss current active matter models that incorporate magnetic forces, addressing pairwise interactions (Sec.~\ref{Sec:3}) and particle shape (Sec.~\ref{Sec:4}), as well as the effects of confinement and external fields (Sec.~\ref{Sec:5}). Our review provides a guided tour of the opportunities and open challenges in magnetic active matter, highlighting both single-particle and collective phenomena, along with promising applications further discussed in the conclusive section (Sec.~\ref{Sec:5}).

\begin{figure*}[t]
    \centering
    \includegraphics[width=0.95\textwidth]{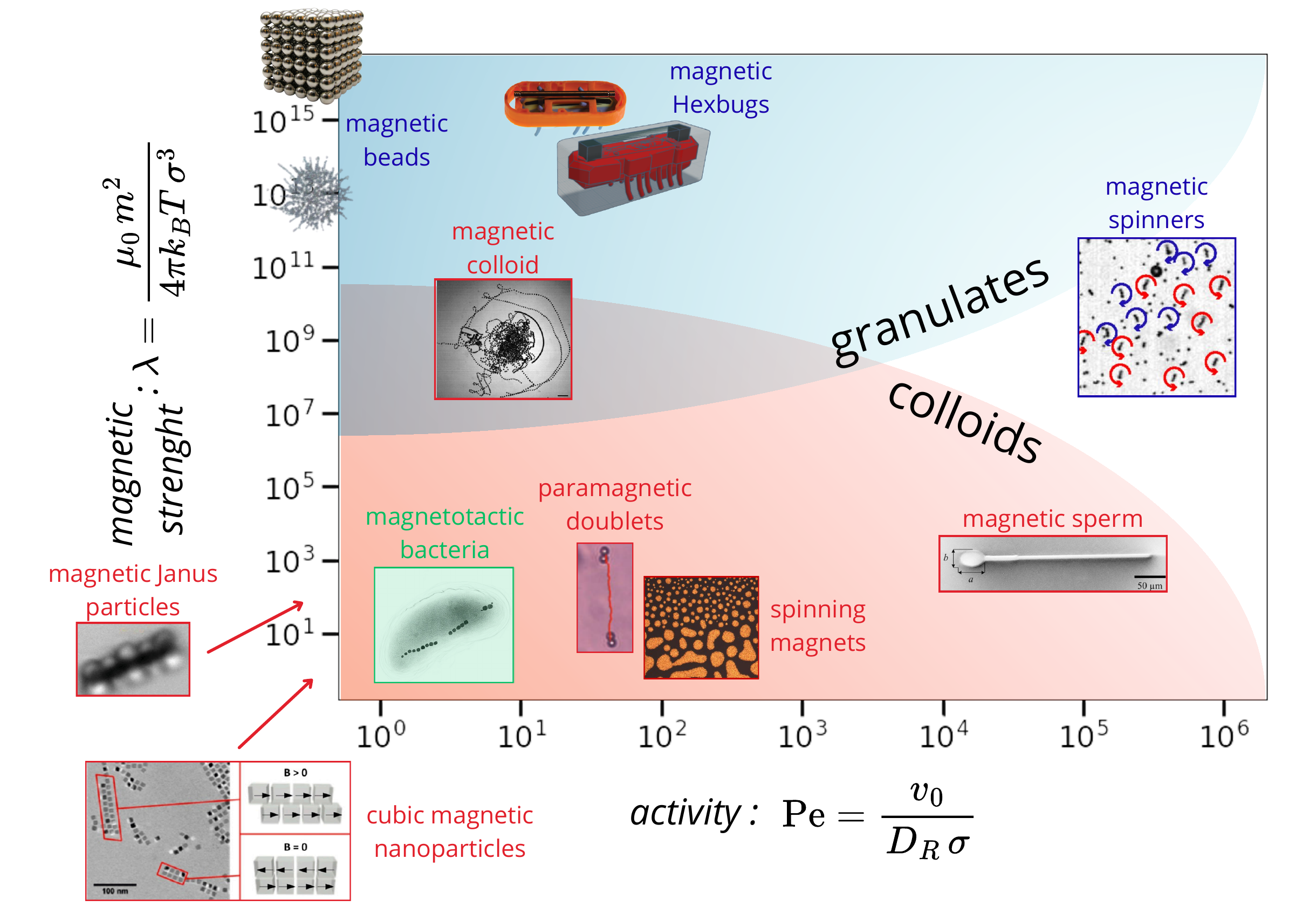}
    \caption{\textit{Schematic overview of magnetic active matter.} The schematic overview reports several experimental realizations of magnetic active matter as a function of their activity, $\text{Pe} = \frac{v_0}{D_R \sigma}$, and the strength of the magnetic interactions, $\lambda = \frac{\mu_0 m^2}{4\pi k_B T \sigma^3}$. The experimental realizations are also classified into colloids~\cite{tierno2012steering, yan2015colloidal, taheri2015selfassembly, magnetosperm2016microrobot, kaiser2017flocking, soni2019oddfreesurface} (red area), granular systems~\cite{gonzalez-gutierrez2013aggregation, schoenke2015infinite, kokot2017active, sepulveda2021bioinspired, musacchio2026fluidization} (blue area) and biological systems \cite{marmol2024magnetotactic} (green area).
}
    \label{fig:fig1}
\end{figure*}

\section{Magnetic Nature in Synthetic and Biological Matter: Dipole Moment}\label{Sec:2}

In nature, diverse animal species evolved internal compasses to detect Earth's magnetic field, many relying on the same principle that researchers independently converged on in the laboratory: embedding a magnetic dipole in a self-propelled agent. Across more than ten orders of magnitude in size (Fig.~\ref{fig:fig1}), this principle gives rise to a class of systems -- magnetic active matter -- in which magnetic alignment, self-propulsion, and collective interactions combine to generate behaviors that neither magnetic nor active matter exhibits in isolation. Representative realizations of magnetic active matter include programmable microswimmers at the microscale, bioconvective suspensions and magnetic granular matter at the macroscale, and swarming robotic collectives at even larger scales.
Beyond their fundamental interest, these systems also enable applications ranging from targeted cancer therapy and biosensing to programmable transport and environmental remediation.

Across all these scales, these systems can be mainly classified and compared depending on their level of activity and strength of their magnetic interactions  (Fig.~\ref{fig:fig1}). The first is quantified by the P\'eclet number which sets the activity strength by measuring the particle persistence length compared to its size; the strength of magnetic interactions is quantified by the magnetic coupling parameter, which compares dipolar interaction energy to thermal energy. We refer to Sec.~\ref{Sec:3} for the definitions of these quantities. 
Remarkably, biological systems, colloidal microswimmers, and granular robots populate distinct but overlapping regions of this parameter space--and the theoretical language developed for one transfers directly to the others. This cross-scale coherence is what makes magnetic active matter a uniquely unified laboratory for nonequilibrium physics.

Below, we list the main experimental realizations of magnetic active matter depending on the length scale, ranging from nanoscale and microscale to macroscale, that characterize them, while these systems are compared depending on their activity value and strength of magnetic interactions in Fig.~\ref{fig:fig1}.

\subsubsection{Nanoscale}

Synthetic iron-oxide nanoparticles magnetite (Fe$_3$O$_4$) and maghemite 
($\gamma$-Fe$_2$O$_3$) illustrate how richly magnetic behavior depends on 
size~\cite{dennis2013physics, karimi2013nano}. Below the single-domain limit ($\approx 50~\mathrm{nm}$ for magnetite), thermal fluctuations overcome the magnetic anisotropy barrier, yielding zero remanence 
in the absence of an external field yet high susceptibility when one is 
applied~\cite{dennis2013physics, dutz2014magnetic}. 
This combination underlies localized tumor hyperthermia, MRI contrast, guided drug 
delivery, and biosensing~\cite{karimi2013nano, dennis2013physics, 
dutz2014magnetic} a toolkit that operates effectively in vivo through surface 
functionalization with biocompatible polymers~\cite{karimi2013nano}.

Nature arrived at these same design principles through evolution. Magnetotactic 
bacteria biomineralize single-domain Fe$_3$O$_4$ crystals--magnetosomes-in 
chains whose collective dipole passively orients the cell along geomagnetic field 
lines~\cite{kirschvink2001magnetite, lin2020origin}. The same physics scales to 
multicellular animals: single-domain magnetite 
provides the orientational torque for navigation in fish, birds, and 
insects~\cite{kirschvink2001magnetite, johnsen2008magnetoreception}. The reach of 
magnetic nanoparticle physics through living matter extends further since human 
brain tissue harbors at least $5\times10^6$ single-domain crystals per gram with 
morphologies resembling bacterial magnetosomes~\cite{kirschvink1992magnetite}, and 
eukaryotic cells can acquire magnetoreception through endosymbiotic 
bacteria~\cite{bolzoni2026magnetoreception}.


\subsubsection{Microscale}

Magnetotactic bacteria exemplify the power of coupling a permanent dipole to 
self-propulsion~\cite{klumpp2019swimming, faivre2008magnetotactic}. A chain of $10$--$30$ magnetosome crystals produces a 
total moment of $\sim10^{-16}$--$10^{-15}~\mathrm{A\,m^2}$; beyond the chain 
length, the cell's magnetic field reduces to that of a single point dipole, 
making magnetotactic bacteria canonical biological realizations of a 
self-propelled magnetic dipole. Under confinement in droplets, shells, or between 
parallel plates, dense populations of these bacteria develop bioconvective 
patterns driven by the interplay between magnetic alignment and 
hydrodynamics~\cite{thery2020selforganisation, vincenti2019magnetotactic, birjukovs2025magnetic}. 

\begin{figure*}[t]
    \centering
    \includegraphics[width=\textwidth]{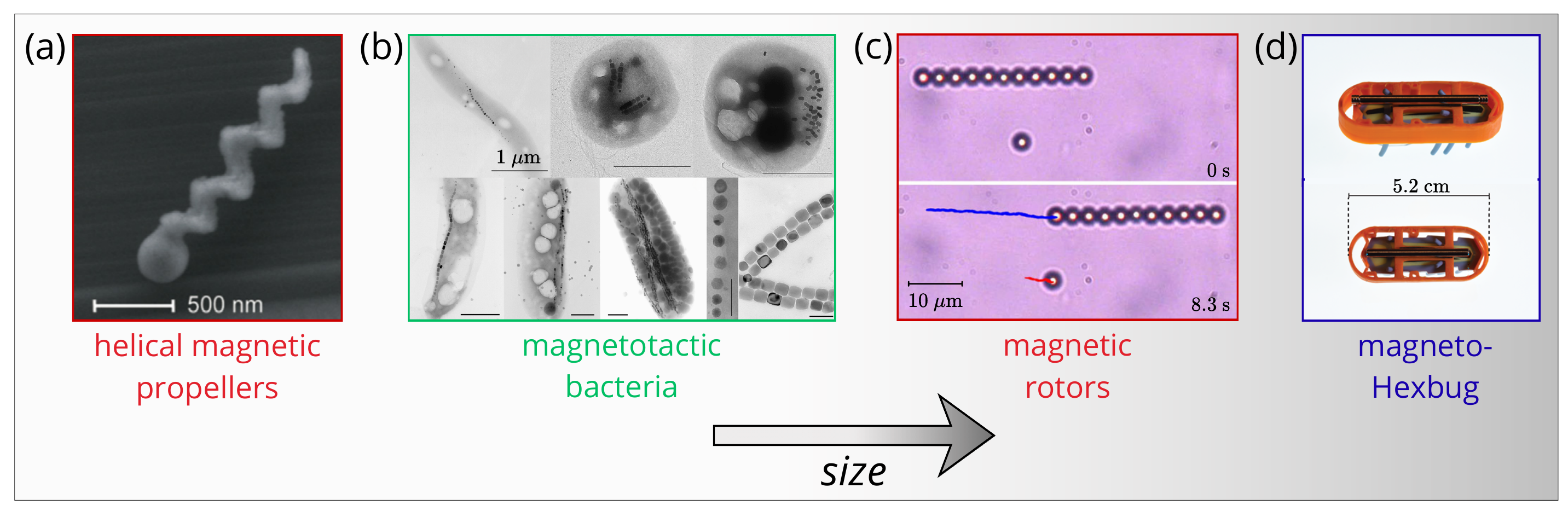}
    \caption{\textit{Active magnetic particles at different scales.} Active magnetic particles at various scales: from (a) nanometer-sized helical magnetic propellers~\cite{klumpp2019swimming}, through (b) micrometer-scale magnetotactic bacteria~\cite{faivre2008magnetotactic} and (c) magnetic rotors~\cite{martinez2015colloidal}, to (d) centimeter-scale magnetic robots~\cite{sepulveda2021bioinspired}.}
    \label{fig:fig2}
\end{figure*}

Synthetic microswimmers in the same size range colloidal particles embedding 
hematite or iron oxide match these moments, and in particle aggregates or 
strongly magnetized colloids dipolar forces become relevant over micrometer-scale 
distances, underpinning field-driven assembly and directed 
propulsion~\cite{tierno2014recent, ghosh2009controlled,Li19}. A compelling example 
is helical magnetic propellers, which mimic bacterial flagella and can be 
steered through viscous media under rotating fields to transport cargo for in vitro 
fertilization~\cite{ghosh2009controlled, magnetosperm2016microrobot, 
klumpp2019swimming}. More broadly, under rotating or oscillating external 
fields these systems exhibit rolling, corkscrew, and flagellar-like propulsion 
modes that can be switched in real time~\cite{martinez2015colloidal, gao2023rolling, xie2019reconfigurable} 
(Fig.~\ref{fig:fig2}), and dense suspensions self-organize into chains, rotors, 
and vortex lattices~\cite{vincenti2019magnetotactic, xie2019reconfigurable, 
martinez2015colloidal, gao2023rolling}. Synthetic surface swimmers can also be designed specifically for the air-liquid interface ('magnetic snakes')~\cite{Snezhko09}. This 
tunability structure and propulsion mode both selectable by field 
parameters illustrates the unique programmability that magnetic active matter offers over purely hydrodynamic microswimmers.

\subsubsection{Macroscale}

At macroscopic scales, where thermal fluctuations are negligible and stochasticity arises from mechanical jitter and substrate inhomogeneities, magnetic self-propelled particles (MSPPs) provide a versatile model platform~\cite{sepulveda2021bioinspired, ledesma2023magnetized,obreque2026dynamics, musacchio2026fluidization} to shed light on the interplay between activity and magnetic interactions. At these scales, self-propelled particles can be designed as vibration driven robots, self-propelling because of an internal motor \cite{arbel2025mechanical}, as in the case of Hexbug particles~\cite{obreque2026dynamics, musacchio2026fluidization}, or as consequence of the vertical vibration of the plate \cite{scholz2018rotating,kumar2014flocking}, as occurs in asymmetric 3d-printed particles often called vibrobots~\cite{antonov2024inertial} or spinners~\cite{scholz2018rotating}.
These centimeter-scale robots can be equipped by neodymium magnets so that they interact through permanent dipole--dipole forces. 
Since oscillating magnetic fields are impractical at these scales, geometric confinement replaces external magnetic fields as the primary control parameter. Curved or polygonal enclosures can stabilize circulating trajectories, promote clustering, and induce symmetry breaking~\cite{musacchio2026fluidization, sepulveda2021bioinspired, obreque2026dynamics}.

Magnetically bound chains of MSPPs extend the biological analogy to flagellar motion: fixing the chain head while activating the remaining particles triggers a spontaneous buckling instability, producing transverse oscillations that mimic flagellar beating through internal propulsion and dipolar cohesion alone~\cite{kiani2015elastic}.

\subsection{Why a Point Dipole?}

\begin{figure}[t]
    \centering
    \includegraphics[width=0.8\columnwidth]{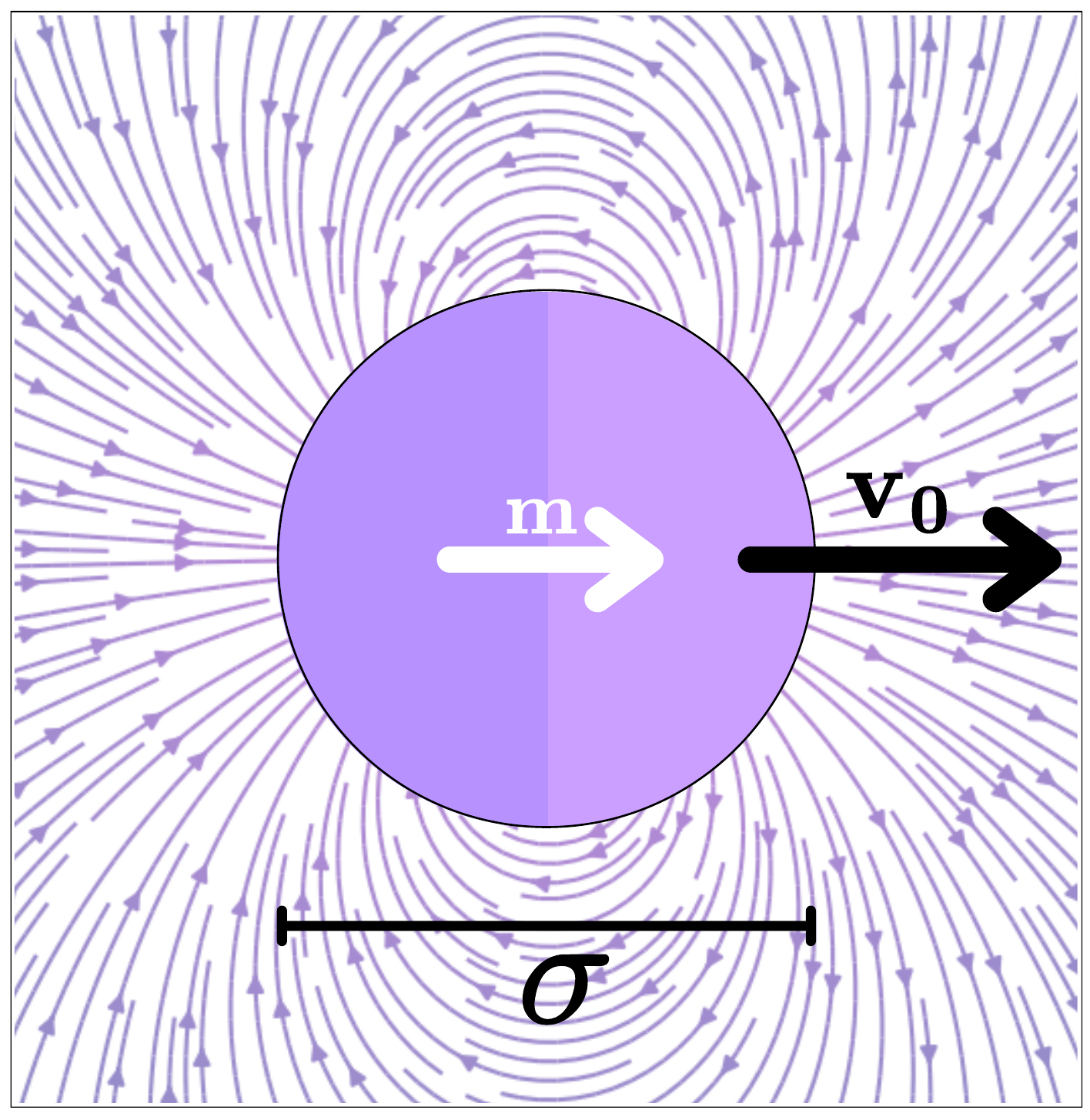}
    \caption{\textit{Dipolar active particle.} Sketch of a dipolar active particle: the cap of the particle indicates its orientation; the black arrow represents its self-propelled velocity, $v_0$, while the white arrow denotes its dipolar moment, \textbf{m}. The violet lines in the background represent the magnetic field lines generated by the dipolar moment.}
    \label{fig:fig3}
\end{figure}

Modeling magnetic active particles as point dipoles (Fig.~\ref{fig:fig3}) 
provides a minimal yet powerful description where the approximation is valid whenever 
the interparticle distance exceeds the size of the magnetic source, reducing the 
field to its leading-order, long-range contribution. The magnetic moment 
$\mathbf{m}$ ($\mathrm{A\,m^2}$) then fully characterizes both the strength and 
orientation of the field the particle generates:
\begin{equation}
    \mathbf{B}(\mathbf{r}) = \frac{\mu_0}{4\pi r^3} 
    \left[ 3(\mathbf{m} \cdot \hat{\mathbf{r}})\hat{\mathbf{r}} - \mathbf{m} \right],
\end{equation}
where $\mu_0 = 4\pi \times 10^{-7}~\mathrm{N\,A^{-2}}$ and $\hat{\mathbf{r}}$ 
is the unit vector from dipole to observation point. The $1/r^3$ decay makes dipolar interactions simultaneously
long-ranged and strongly anisotropic and, unlike their
electrostatic counterpart, unscreened -- the physical origin of the
rich self-assembly described in subsequent sections.

Aside from being physically compact and easier to handle both analytically and in simulation, for many physical configurations this is indeed an accurate approximation. We will discuss the limitations in slightly more detail here. Firstly, as mentioned above, the interparticle distance criterion is often satisfied by virtual of the particle geometry. At a smaller scale, colloidal magnetic particles require stabilization via some polymeric shell or coating, while at a larger scale, the magnetic component of biological or synthetic microbots is commonly at the center of the particle. Regarding the magnetization at smaller scales, there are two additional simplifications: firstly, the particle material (and geometry) will determine the actual magnetization. For instance, magnetic cuboidal particles~\cite{Blon17} can exhibit multiple different ground states. Additionally, magnetically capped Janus particles~\cite{Kretzschmar11} exhibit self-assembly that is not well-described by central point dipoles. In these cases, modeling the magnetization as a point dipole shifted away from the particle center has been proven to agree with experimental data~\cite{Kretzschmar19,Ubaldo20}, or in more complex cases, distributions of non-collinear shifted dipoles~\cite{Steinbach19}, although this approach is limited for a large magnetization distribution~\cite{Gemming16}. Such distributions and offsets of particles can also be extended to model higher-order multipoles, such as quadrupoles~\cite{Rosenberg26}. Secondly, another simplification is the assumption of hard magnetic materials. Under the influence of a magnetic field, magnetic particles physically have multiple different modes of relaxation. The point dipole model only allows for Brownian relaxation, although extensions to include N{'}eel relaxation, specifically the Thermal Stoner Wolfarth model, are gaining traction as increasing computational power renders them feasible~\cite{Mostarac25}. Magnetodynamic simulations have also been carried out to understand driven nanoparticle assembly~\cite{Bui2025}. Finally, it must also be noted that approximating many-particle dipolar interactions as a superposition of pairwise interactions is a modeling assumption, which does not hold fully for many-body soft magnetic systems~\cite{romeis2026}.

The point-dipole approximation is well validated across scales. For magnetotactic bacteria, 
typical cell separations exceed the magnetosome chain length, so the dipolar 
component dominates at biologically relevant distances~\cite{klumpp2019swimming}. 
For synthetic colloids and field-driven assemblies, the overdamped dynamics are 
governed by the balance between magnetic torques, hydrodynamic interactions, 
self-propulsion, and rotational diffusion~\cite{guzman2016fission, 
martinez2015colloidal, parage2025modulation, 
telezki2020simulations, gao2023rolling, xie2019reconfigurable}. At the macroscale, 
MSPPs carry neodymium magnets with moments $m\sim10^{-2}$--$1~\mathrm{A\,m^2}$, 
and the point-dipole model accurately describes magnetic orbiting, chain 
buckling, and collective dynamics as long as interparticle distances exceed the 
magnet size~\cite{obreque2026dynamics, musacchio2026fluidization, 
kiani2015elastic, gao2025soft}. When this condition is violated--when the 
magnet length becomes comparable to the separation--a dumbbell model 
representing the magnet as two opposite magnetic charges at finite separation 
captures near-field corrections more accurately~\cite{sepulveda2021bioinspired}.

\section{Magnetic Self-propelled Particles}\label{Sec:3}

We consider a system of $N$ magnetic self-propelled particles (MSPPs) moving in 
two dimensions in a fluid medium with viscosity $\mu$. Each particle is 
characterized by its position $\mathbf{r}_i(t)$ and unit orientation vector 
$\hat{\mathbf{n}}_i(t) = [\cos\theta_i, \sin\theta_i]$, which defines both its 
propulsion direction and the axis of its magnetic dipole moment 
$\mathbf{m}_i = m\hat{\mathbf{n}}_i$.
In the low Reynolds number regime relevant to microscale systems, inertial 
effects are negligible and the dynamics are governed by the overdamped Langevin 
equations:
\begin{subequations}
\label{eq:micro}
\begin{flalign}    
\dot{\mathbf{r}}_i &= v_0 \hat{\mathbf{n}}_i 
- \frac{\nabla_{\mathbf{r}_i}}{\gamma_T}  \left( \sum_{j \neq i} 
U^{\text{EV, D}}_{ij} + U^{\text{EF}}_i \right) 
+ \mathbf{u}_i^{\text{HI}} + \boldsymbol{\xi}_{i,T}
\label{eq:micro_trans} \\
\dot{\hat{\mathbf{n}}}_i &= \left( \boldsymbol{\xi}_{i,R}(t) 
- \mathbf{T}_i^{\text{EV}} - \mathbf{T}_i^{D} - \mathbf{T}_i^{EF} -\mathbf{T}_i^{HI}\right)\times \frac{\hat{\mathbf{n}}_i}{\gamma_R} , 
\label{eq:micro_rot}
\end{flalign}
\end{subequations}

\noindent where $\boldsymbol{\xi}_{i,T}(t)$ and $\boldsymbol{\xi}_{i,R}(t)$ are Gaussian white noise terms with zero mean, satisfying $\langle \boldsymbol{\xi}_{i,T}(t)\cdot\boldsymbol{\xi}_{i,T}(t') \rangle = 2 d D_T\delta(t-t')$ and $\langle \xi_{i,R}(t)\xi_{i,R}(t') \rangle = 2D_R\delta(t-t')$. Here, $D_T$ and $D_R$ denote the translational and rotational diffusion coefficients, respectively, while $d$ is the dimensionality of the system. Finally, $\gamma_T$ and $\gamma_R$ are the spatial and rotational friction coefficients, respectively, and $v_0$ is the self-propulsion speed, i.e., the particle's activity.
The inverse rotational diffusion coefficient, $1/D_R$, defines the characteristic persistence time of an active particle's trajectory and, consequently, its persistence length, $v_0/D_R$, namely the average distance traveled before its direction changes appreciably. Following previous work, we quantify the activity strength by introducing the P\'eclet number,
\begin{equation}
\mathrm{Pe}=\frac{v_0}{D_R\sigma},
\end{equation}
which compares the persistence length to the particle diameter $\sigma$. Throughout this work, and in Fig.~\ref{fig:fig1}, $\mathrm{Pe}$ is used to characterize the activity of different experimental realizations of active matter systems.

While $D_T$ originates from the random collisions with the solvent molecules, we remark that only in active colloidal systems is the rotational diffusion coefficient $D_R$ related to the environmental temperature through the Einstein relation $D_R = 2k_B T/\gamma_R$, where $k_B$ is the Boltzmann constant. In contrast, in most active matter systems, including biological microorganisms, $D_R$ is entirely unrelated to the environmental temperature and is instead determined by other physical mechanisms.
Conservative interactions enter through pairwise potentials $U_{ij}^{\alpha}$ 
and associated torques $\mathbf{T}_i^{\alpha} = \hat{\mathbf{n}}_i \times 
\nabla_{\hat{\mathbf{n}}_i} U^{\alpha}$, with $\alpha \in \{\text{EV}, \text{D}\}$ 
denoting excluded volume and dipolar interactions, respectively. The 
external-field torque is given explicitly by 
$\mathbf{T}_i^{\text{EF}} = \mu_0 m\,\hat{\mathbf{n}}_i \times \mathbf{H}$, 
while $\mathbf{T}_i^{\text{HI}}$ and $\mathbf{u}_i^{\text{HI}}$ are torque arising from hydrodynamic interactions.

Short-range excluded volume is modeled by the repulsive WCA potential for 
disk-shaped particles, or by Gay--Berne and Yukawa-like potentials for 
anisotropic shapes such as ellipsoids or rods. Long-range magnetic coupling 
between point dipoles is given by
\begin{equation}
U^{\text{D}}_{ij} = \frac{\mu_0 m^2}{4\pi r_{ij}^3} \left[ \hat{\mathbf{n}}_i 
\cdot \hat{\mathbf{n}}_j - 3\frac{(\hat{\mathbf{n}}_i \cdot \mathbf{r}_{ij})
(\hat{\mathbf{n}}_j \cdot \mathbf{r}_{ij})}{r_{ij}^2} \right],
\label{dipoleinter}
\end{equation}
where $\mathbf{r}_{ij} = \mathbf{r}_j - \mathbf{r}_i$. The modeling of magnetic particle shapes is discussed in more detail in Ref~\cite{Domingos26}.

Expressing the interparticle distance $\mathbf{r}_{ij}$ in units of the particle diameter $\sigma$, the prefactor of $U^{\text{D}}_{ij}$ defines the dimensionless magnetic interaction strength
\begin{equation}
\lambda = \frac{\mu_0 m^2}{4\pi \sigma^3 k_B T},
\end{equation}
where $T$ is the environmental temperature and $k_B$ is the
Boltzmann constant. The translational diffusion coefficient is
related to the temperature through the Einstein relation,
$D_T = 2k_B T / \gamma_T$. The parameter $\lambda$ is used in Fig.~\ref{fig:fig1} to compare the magnetic interaction strength of different magnetic active systems.

\begin{figure}[t]
    \centering
    \includegraphics[width=\columnwidth]{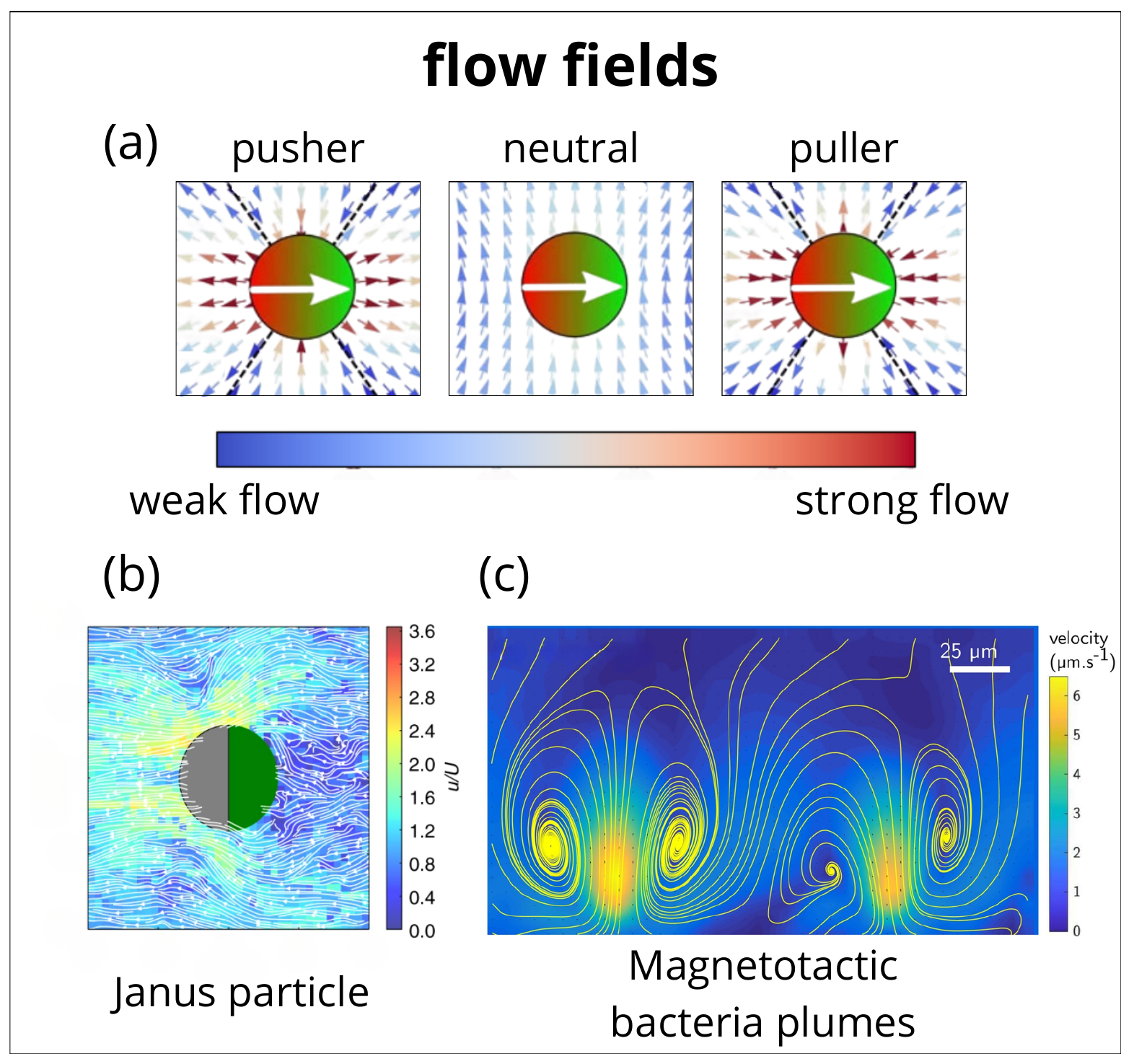}
    \caption{\textit{Hydrodynamic flow fields.} Hydrodynamic flow fields (color gradient) produced by (a) pusher, neutral, and puller particles~\cite{guzman2016fission}, (b) a Janus particle~\cite{campbell2019experimental}, and (c) magnetotactic bacteria plumes~\cite{thery2020selforganisation}.}
    \label{fig:fig4}
\end{figure}

At low Reynolds number, the far-field flow of a self-propelled particle is well 
approximated by a force dipole (Fig.~\ref{fig:fig4}(a)):
\begin{equation}
\mathbf{u}(\mathbf{r}) = \frac{p}{8\pi\mu r^2} \left[ 3(\hat{\mathbf{n}} \cdot 
\hat{\mathbf{r}})\hat{\mathbf{r}} - \hat{\mathbf{n}} \right],
\label{eq:stokesdipole}
\end{equation}
 where $\mu$ is the dynamic viscosity and the sign of the coefficient $p$ determines whether the swimmer is a pusher ($p>0$), a puller ($p<0$), or a neutral swimmer ($p=0$)~\cite{guzman2016fission, campbell2019experimental}.
For MSPPs, magnetic forces and torques generate additional hydrodynamic contributions, so that the total flow experienced by particle $i$ due to all others is:
\begin{equation}
\mathbf{u}_i^{\text{HI}} = \sum_{j \ne i} \left[ \mathbf{G}(\mathbf{r}_{ij}) 
\mathbf{f}^j + \left( \frac{\mathbf{r}_{ij}}{8\pi\mu r_{ij}^3} \times 
\mathbf{T}_j^{\text{D,EF}} \right) + u_j(\mathbf{r}_{ij}) \right],
\label{uhi}
\end{equation}
where the first term is the Stokeslet from the magnetic dipolar force 
$\mathbf{f}_j = -\nabla_{\mathbf{r}_j}U^{\text{D}}$ propagated via the Oseen 
tensor $\mathbf{G}(\mathbf{r})$, and the second is the rotlet from magnetic 
torques $\mathbf{T}_j^{\text{D,EF}}$. Rotlets generate circulating flows that 
drive long-range orientational entrainment, particularly relevant in rotating 
chain assemblies~\cite{tierno2014recent, thery2020selforganisation}. The 
corresponding hydrodynamic torque on particle $i$ is:
\begin{equation}
\mathbf{T}_i^{\text{HI}} = \sum_{j \ne i} \left( \frac{\mathbf{r}_{ij}}{r_{ij}^3} 
\times \frac{\mathbf{f}_j}{8\pi\mu} \right).
\label{thi}
\end{equation}
Experimental observations (Fig.~\ref{fig:fig4}(b),(c)) confirm that these 
fluid-mediated couplings can sustain collective motion even when direct dipolar 
interactions are suppressed~\cite{tierno2014recent,thery2020selforganisation, 
gao2025soft}.

\subsection{Macroscale: Inertial Particles with Overdamped Rotation}

Macroscopic realizations of magnetic active matter, such as systems based on 
Hexbug robots or similar centimeter-scale platforms, operate in a fundamentally 
different regime from their microscopic counterparts. These systems belong to 
the class of \emph{dry active matter}, where motion takes place on solid 
substrates and interactions are mediated by direct mechanical contact and 
magnetic forces, rather than by fluid-mediated hydrodynamics.

A key distinction of macroscopic systems is that translational inertia cannot 
be neglected. The finite mass of the particles introduces inertial effects in 
their displacement, leading to delayed responses, transient oscillations, and 
underdamped motion in confined or interacting configurations. In contrast, 
rotational dynamics remain strongly overdamped due to the large frictional 
torques generated by substrate contact and internal dissipation. This separation 
of timescales justifies a modeling approach in which particle positions follow 
inertial dynamics while orientations evolve overdampedly, as validated 
experimentally and theoretically in recent studies of macroscopic active 
matter~\cite{baconnier2025self}.

The equations of motion for MSPPs in the dry macroscopic regime are:
\begin{subequations}
\begin{eqnarray}
M \ddot{\mathbf{r}}_i &=& F_0 \hat{\mathbf{n}}_i 
- \gamma_T \dot{\mathbf{r}}_i \nonumber \\ 
&&- \nabla_{\mathbf{r}_i} \left( \sum_{j \neq i} \left(U^{\text{EV}}_{ij} 
+ U^{\text{D}}_{ij} \right) + U^{\text{EF}}_i \right), 
\label{eq:macro_trans}\\
\dot{\hat{\mathbf{n}}}_i &=& \frac{1}{\gamma_R} \left( \boldsymbol{\xi}_{i,R}(t) 
+ \beta\, \hat{\mathbf{n}}_i \times \hat{\mathbf{v}}_i \right. \nonumber \\ 
&&\left.- \mathbf{T}_i^{\text{EV}} 
- \mathbf{T}_i^{\text{D}} 
- \mathbf{T}_i^{\text{EF}} \right) \times \hat{\mathbf{n}}_i + \boldsymbol{\xi}_{i,T}.
\label{eq:macro_rot}
\end{eqnarray}
\end{subequations}
Here, $M$ is the particle mass, while $F_0$ denotes the active force generated by the internal motor, which as before provides a self-propelled velocity $F_0= \gamma_T v_0$. 
As in Eqs.~\eqref{eq:micro}, $\gamma_T$ and $\gamma_R$ represent the translational and rotational friction coefficients, while 
$\boldsymbol{\xi}_{i,T}(t)$ and $\boldsymbol{\xi}_{i,R}(t)$ are Gaussian white noises such that $\langle \boldsymbol{\xi}_{i,T}(t)\cdot\boldsymbol{\xi}_{i,T}(t') \rangle = 2dD_T\delta(t-t')$ and $\langle \xi_{i,R}(t)\xi_{i,R}(t') \rangle = 2D_R\delta(t-t')$. These term denote stochastic translational and angular fluctuations generated by 
mechanical noise and substrate heterogeneities, playing a role analogous to 
rotational diffusion in microscopic systems.
As before, The potentials $U^{\text{EV}}$ and $U^{\text{D}}$ represent excluded volume and magnetic dipole--dipole interactions, respectively, $U^{\text{EF}}$ accounts for external fields or confinement.

In addition to inertial effects~\cite{antonov2024inertial}, macroscopic active matter is typically characterized by a \emph{self-alignment torque}~\cite{baconnier2025self} governing the dynamics of orientation vector $\hat{\mathbf{n}}_i$. This torque is proportional to $\beta\hat{\mathbf{n}}_i \times \mathbf{v}_i$, and tends to align the particle orientation with its velocity with strength $\beta$.
This term does not originate from hydrodynamic coupling or thermal fluctuations, but instead arises naturally when the geometric center of the particle does not coincide with its center of mass~\cite{baconnier2025self}. This mechanism has been experimentally demonstrated in active granular systems and has been proposed to underlie the behavior observed in several experiments on living cells.
Self-alignment has been shown to play a central role in inducing collective motion~\cite{baconnier2025self}, inducing chiral trajectories, controlling large-scale organization in robotic active 
matter~\cite{casiulis2025geometric}, as well as leading to flocking behavior in dense systems or flocking clusters at intermediate density~\cite{musacchio2025self}. When external magnetic fields are present, $\mathbf{T}_i^{\text{EF}}$ provides an additional alignment torque, though in many macroscopic platforms control is achieved primarily through geometry and confinement rather than applied fields.

Overall, the dry macroscopic regime provides a minimal yet powerful framework 
for studying magnetic active matter, where inertia, self-alignment, and dipolar 
interactions combine to produce rich nonequilibrium dynamics complementary to 
those of microscale swimmers--highlighting how similar interaction rules give 
rise to qualitatively different behaviors when noise, dissipation, and inertia 
are reorganized across scales.

\section{Particle Shape}\label{Sec:4}

As has been alluded to in multiple previous subsections, the particle shape is a crucial parameter to tailor the system behavior~\cite{guzman2026collective}. This is true both for the magnetic core, from magnetization of the particle to the interparticle minimum energy configuration, and for the overall shape of the aggregate, both from the perspective of a pure active context (\textit{e.g.} shape-induced suppression of motility-induced phase separation~\cite{liao2020dynamical}) or the more complex propulsion and hydrodynamics of anisotropic swimmers. 

\begin{figure}[t]
    \centering
    \includegraphics[width=\columnwidth]{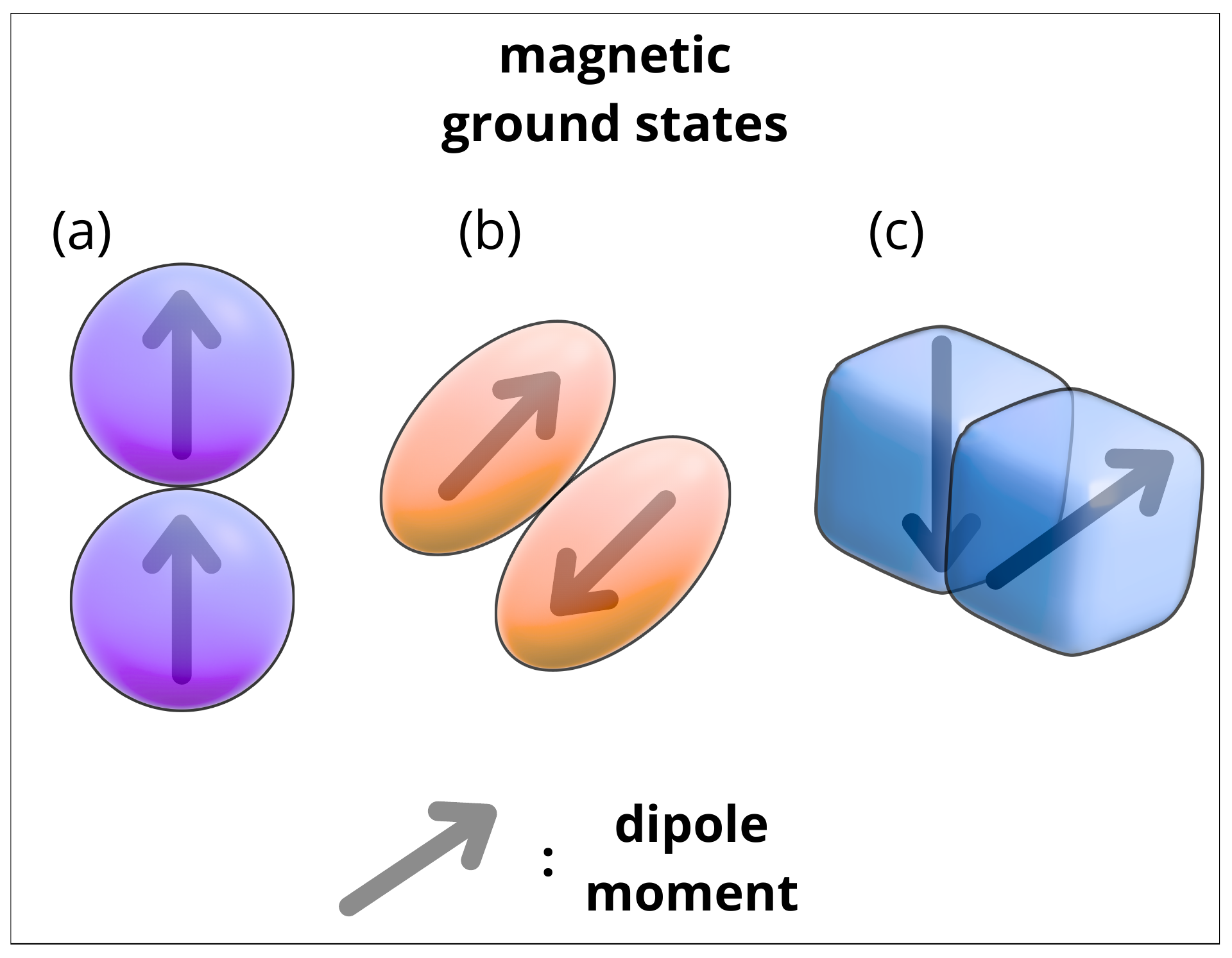}
    \caption{\textit{Magnetic ground state of particles with different shapes.} Magnetic ground state for (a) spherical particles, (b) ellipsoidal particles, and (c) cubic particles.}
    \label{fig:fig5}
\end{figure}

Pairwise dipole-dipole interactions determine the equilibrium distance and 
angular configuration between active particles depending on particle 
shape~\cite{soni2019oddfreesurface, kantorovich2013influence, tierno2014recent, 
vanesse2023collective}. For a spherical shape, excluding the anisometry, the ground state of a pairwise interaction is in the head-to-tail orientation shown in Fig.~\ref{fig:fig5} a), which is also easy to see from the interaction potential~\eqref{dipoleinter}. This minimum is equally the origin of the characteristic interaction strength parameter $\lambda$ (as seen in Fig.~\ref{fig:fig1}), as the energy of two head-to-tail aligned particles at close contact $d$ is then precisely $- 2 \lambda$ in units of $k_B T$. In a passive equilibrium system, we would therefore expect the low-density ground state of the system to be a cluster-gas of chains and rings~\cite{Sciortino13}. These are also found in active magnetic microswimmers ~\cite{guzman2016fission,obreque2026dynamics,telezki2020simulations}, although due to the activity introducing additional perturbation to the system, metastable defect structures such as y-junctions are already found in isolated small-particle aggregates~\cite{kaiser2015active}. 

Self-propulsion introduces 
persistent driving forces that compete with these equilibrium 
configurations~\cite{telezki2020simulations,kelidou24,Royall25}; at low 
motility, dipolar active Brownian particles aggregate into chains, whereas 
increasing activity breaks these structures, driving a transition to an 
isotropic fluid or, at high densities, to a flocking state in which dipolar 
coupling suppresses motility-induced phase separation~\cite{liao2020dynamical, 
kokot2017active}. For active particle ensembles, we see a preponderation of  filamentary 
structures~\cite{kogler2015lane}, such as in ~\cite{kokot2015emergence}, 
 where suspensions of magnetic microparticles under 
alternating fields self-assemble into single-particle-thick wires and dynamic 
spinner arrays, and arrays of magnetic rotors adopt 
collective modes governed by the competition between dipolar coupling and 
external field strength~\cite{massana2019tunable}. In macroscopic vibrated systems, the ratio of kinetic to 
magnetic potential energy governs transitions between chains, vortices, flocks, 
and stripe patterns~\cite{kaiser2017flocking, han2020emergence}, and dipolar 
coupling lowers the critical density threshold for collective 
behavior~\cite{vanesse2023collective}. Chiral active particles with dipolar 
interactions further enrich this landscape, generating vortex formation and 
phase separation even below the motility-induced phase separation density threshold~\cite{liao2021emergent}.

As illustrated by Fig.~\ref{fig:fig5} b) and c), this picture is further complicated by particle shape. As a simple approximation, magnetotactic bacteria are often represented by rods~\cite{klumpp2019swimming}, for which the minimal energy configuration depends on the aspect ratio~\cite{kantorovich2013influence}. Similarly, cuboidal magnetic particles have been thoroughly investigated due to their promise for hyperthermia applications~\cite{Poon25}; and as the magnetization will depend on the material, the resulting structure formations can vary from rings to zig-zag structures~\cite{Donaldson18}. In an active context, this affects both the steering via external fields and the hydrodynamics of particle transport in a microchannel~\cite{Martin22}, especially if the activity and magnetic moment are not coaligned~\cite{Martin20a}. Moreover, hierarchical aggregates of multiple particles, which can be beneficial for transport problems, unlock additional shapes such as asters~\cite{Snezhko2011}. Similarly, placing charged colloids in a bath or of active magnetic Janus particles gives rise to rotating brush superstructures~\cite{Harder18}.

\section{The Role of Confinement and External Fields}\label{Sec:5}

Geometric confinement reshapes the effective potential landscape experienced 
by MSPPs, enabling control over collective behavior without time-varying 
external fields~\cite{Lowen18}. Parabolic or curved boundaries impose restoring forces that 
stabilize circulating trajectories and suppress diffusive 
spreading~\cite{obreque2026dynamics, musacchio2026fluidization, 
sepulveda2021bioinspired}. Polygonal and circular enclosures additionally 
induce symmetry breaking and cluster formation~\cite{musacchio2026fluidization}, 
acting as a geometric analog of external fields in systems where traditional 
control parameters are limited or inaccessible.

External magnetic fields offer direct, real-time control over both individual 
particles and collective behavior in magnetic active matter. At the 
single-particle level, field-responsive microparticles near surfaces can be 
remotely spun, steered, and assembled through controlled field modulations, 
enabling directed transport and assembly with micrometer-scale 
precision~\cite{yan2015colloidal, tierno2014recent,Elschner2024,Han24a}. Alternating or rotating fields applied to dense suspensions of ferromagnetic rollers drive 
global rotation and flocking through spontaneous symmetry breaking between 
clockwise and counterclockwise modes, with orientational noise from shape 
imperfections providing the stochasticity necessary for collective 
order~\cite{kaiser2017flocking}. Concentrated rollers further self-organize into 
multivortex states in unconfined environments, with neighboring vortices 
adopting opposite rotation senses~\cite{han2020emergence}, while rotor arrays 
under competing dipolar coupling and field strength transition between 
alternating-stripe patterns reflecting the spin-ice symmetry of their static 
configuration and field-dominated regimes supporting spatially differentiated 
collective modes. More broadly, magnetic rollers exhibit a rich phase diagram~\cite{Wang19}, including controllable multi-vortex states~\cite{Han24a} and chiral states~\cite{Han21b}.

For active dipolar particles, the interplay between self-propulsion and external 
alignment produces a hierarchy of nonequilibrium structures. Weak fields promote 
disordered chains and percolated networks through competition between 
activity-driven bond breaking and field-promoted bond formation, while strong 
fields collapse these into polarized columnar clusters whose spacing decreases 
monotonically with field intensity~\cite{parage2025modulation, 
telezki2020simulations}. Moderate activity extends the field range over which 
percolated networks persist, and the polarization response crosses from 
super-Langevin to sub-Langevin behavior with increasing 
activity~\cite{parage2025modulation}. At the continuum level, competing 
hydrodynamic and magnetic torques destabilize homogeneous polar states, 
generating bend-twist-driven traveling sheets and dynamical aggregates for 
pusher suspensions, and splay-dominated migratory pillars for pullers, with 
reentrant stability recovered at very strong fields~\cite{koessel2020emergent}. 
Analogous field-driven alignment in magnetotactic bacteria confined in droplets, 
shells, and between plates produces bioconvective patterns through the same 
competition between magnetic alignment and hydrodynamic 
instabilities~\cite{thery2020selforganisation, vincenti2019magnetotactic, birjukovs2025magnetic}.

These principles extend to soft robotic platforms, where field-induced shape 
transformations in magnetic hydrogels from helical to planar 
configurations enable remote sensing and collective signal 
amplification~\cite{gao2025soft}, illustrating how external field control of 
active dipolar systems continues to inspire functional design at the 
intersection of active matter and soft robotics.

\section{Conclusions}\label{Sec:6}
This review has shown that the magnetic dipole moment provides a surprisingly 
unifying thread across twelve orders of magnitude in length. Whether biomineralized 
in a magnetotactic bacterium, embedded in a colloidal microswimmer, or encased 
in a centimeter-scale robot, the same anisotropic $1/r^3$ interaction governs 
the competition between chain formation, ring closure, and dynamic collective motion. 
The P\'eclet number and the magnetic coupling parameter organize 
this phenomenology into a single parameter space (Fig.~\ref{fig:fig1}) that 
encompasses biological, colloidal, and granular realizations a cross-scale coherence that distinguishes magnetic active matter from most other classes of active systems.

The dialogue between biological and synthetic realizations has been productive in both directions. Biology first established the core design principles through single-domain magnetite near the superparamagnetic threshold, stabilized in chains by cytoskeletal filaments, producing a body-fixed dipole whose coupling 
to self-propulsion enables passive geomagnetic 
navigation~\cite{kirschvink2001magnetite, lin2020origin, kiani2015elastic}. Synthetic systems have reproduced and extended these principles, from helical magnetic propellers that mimic bacterial flagella for cargo transport and 
assisted fertilization~\cite{ghosh2009controlled, magnetosperm2016microrobot}, to field-driven colloidal rollers that flock and form multivortex states~\cite{kaiser2017flocking, han2020emergence}, to macroscopic robots whose 
magnetically bound chains spontaneously beat like eukaryotic 
flagella~\cite{kiani2015elastic}. In return, tools developed for synthetic active matter have illuminated magnetotaxis, bioconvection, and the mechanical role of cytoskeletal architecture in ways that purely biological approaches could not.

The theoretical framework developed here, ranging from overdamped and inertial Langevin dynamics, Stokeslet and rotlet hydrodynamics, point-dipole to dumbbell interaction models and continuum instability analysis, is predictive beyond the systems already studied. It identifies the conditions under which motility-induced phase separation is 
suppressed by dipolar coupling~\cite{liao2020dynamical}, the field strengths 
at which polarization crosses from super- to sub-Langevin 
behavior~\cite{parage2025modulation}, and the activity thresholds at which homogeneous polar states destabilize into traveling sheets or migratory pillars~\cite{koessel2020emergent}. Extending these predictions to three dimensions, to mixtures of pushers and pullers with permanent moments, and to regimes where wet and dry dynamics coexist remains an open and tractable challenge.

Several questions stand out as particularly pressing. The functional role of single-domain magnetite in the human brain remains unknown~\cite{kirschvink1992magnetite}, and the recent discovery that entire eukaryotic cells can acquire magnetoreception through 
endosymbiosis~\cite{bolzoni2026magnetoreception} suggests that the biological inventory of magnetic active matter is far from closed. At the synthetic frontier, shape anisotropy beyond the disk geometry rods, helices, asymmetric dumbbells opens new axes of control over ground-state structures and propulsion modes~\cite{kantorovich2013influence, vanesse2023collective, 
tierno2014recent}, while soft magnetic microrobots integrating shape morphing with wireless communication point toward active matter systems capable of autonomous sensing and collective decision-making~\cite{gao2025soft}. Magnetic active matter thus sits at a productive intersection, grounded in well-understood dipolar physics, enriched by biological inspiration, and open toward programmable 
materials, biomedical microrobotics, and the fundamental non-equilibrium physics of self-organizing matter.

\section*{Acknowledgements}
F.G.-L acknowledges Fondecyt Regular 1250913. M.R. acknowledges support from the Alexander von Humboldt-Foundation. 

\vspace{0.1cm}

\section*{Author contributions}
 All authors contributed equally writing the manuscript.

\section*{Additional information}
There are no conflicts to declare.

\bibliographystyle{elsarticle-num}

\bibliography{Bibliography3}

\appendix

\end{document}